\documentclass[prb,twocolumn,showpacs,preprintnumbers,amsmath,amssymb,floatfix]{revtex4}
\usepackage{graphicx}% Include figure files
\usepackage{float}
\usepackage{trfsigns}

\newcommand \be{\begin{eqnarray}}
\newcommand \ee{\end{eqnarray}}

\newcommand \ba{\begin{align}}
\newcommand \eea{\end{align}}
\newcommand {\ket}[1]{|#1\rangle}
\newcommand {\bra}[1]{\langle #1|}

\begin{document}
%\twocolumn[\hsize\textwidth\columnwidth\hsize
           \csname @twocolumnfalse\endcsname
\title{Quantum currents and pair correlation of electrons in a chain of localized dots}
\author{K. Morawetz$^{1,2,3}$ 
%in discussion with\\
%J. Kailasvuori$^3$, O. Bozat$^{2}$, H. Schmidt$^4$  
%and P. Lipavsk\'y$^{4,5}$,
}
\affiliation{$^1$M\"unster University of Applied Sciences,
Stegerwaldstrasse 39, 48565 Steinfurt, Germany}
\affiliation{$^2$International Institute of Physics (IIP),
Av. Odilon Gomes de Lima 1722, 59078-400 Natal, Brazil
%International Center of Condensed Matter Physics, University of Bras\'ilia, 70904-970, Bras\'ilia-DF, Brazil
}
\affiliation{$^{3}$ Max-Planck-Institute for the Physics of Complex Systems, 01187 Dresden, Germany
}
%\affiliation{$^4$Institute of Ion Beam Physics and Materials Research, Helmholtz-Zentrum Dresden-Rossendorf, P.O.Box 51 01 19, 01314 Dresden, Germany}
%\affiliation{$^4$Faculty of Mathematics and Physics,
%Charles University, Ke Karlovu 3, 12116 Prague 2, Czech Republic}
%\affiliation{$^5$Institute of Physics, Academy of Sciences,
%Cukrovarnick\'a 10, 16253 Prague 6, Czech Republic}
\begin{abstract}
The quantum transport of electrons in a wire of localized dots by hopping,
interaction and dissipation is calculated and a representation by an
equivalent RCL circuit is found. The exact solution for the electric-field
induced currents allows to discuss the role of virtual currents to decay
initial correlations and Bloch oscillations. The dynamical response function
in random phase approximation (RPA) is calculated analytically with the help of which the static structure function and pair correlation function are determined. The pair correlation function contains a form factor from the Brillouin zone and a structure factor caused by the localized dots in the wire.
\end{abstract}
\pacs{
%03.65.Nk %      Scattering theory
%,21.45.+v %     Few-body systems
%,72.10.Fk %     Scattering by point defects, dislocations, surfaces, and other imperfections (including Kondo effect)
%,03.65.Ge %     Solutions of wave equations: bound states
%,34.80.Pa %     Coherence and correlation in electron scattering
%,34.10.+x %     General theories and models of atomic and molecular collisions and interactions (including statistical theories, transition state, stochastic and trajectory models, etc.)
%,68.65.Hb %Quantum dots
%,73.22.-f %     Electronic structure of nanoscale materials: clusters, nanoparticles, nanotubes, and nanocrystals
%,79.20.Rf %    Atomic, molecular, and ion beam impact and interactions with surfaces
%, 61.14.Dc %    Theories of diffraction and scattering
%,61.46.+w %     Nanoscale materials: clusters, nanoparticles, nanotubes, and nanocrystals 
71.45.Gm, %      Exchange, correlation, dielectric and magnetic response functions, plasmons
78.20.-e, %      Optical properties of bulk materials and thin films 
78.47.+p, %      Time-resolved optical spectroscopies and other ultrafast optical measurements in condensed matter 
42.65.Re, %     Ultrafast processes; optical pulse generation and pulse compression
82.53.Mj, %     Femtosecond probing of semiconductor nanostructures 
71.10.+w, %Theories and models of many-electron systems
 71.70.Ej, %Spin-orbit coupling, Zeeman and Stark splitting, Jahn-Teller effect 
75.76.+j, %Spin transport effects
 85.75.Ss %Magnetic field sensors using spin polarized transport 
% 03.75.-b, %Matter waves
%74.20.-z, %Theories and models of superconducting state
%67.85.-d, %Ultracold gases, trapped gases
%64.70.Nd, %Structural transitions in nanoscale materials
%75.70.Ak, %Magnetic properties of monolayers and thin films
%75.30.Fv, %Spin-density waves 
}
\maketitle
%    \vskip2pc]

\section{Introduction}

The physics and the industrial usage of chains of molecules has been a subject
of immense activity during last years \cite{XLZM03,Lec07,WHW09,B14}. Let us mention only some selected examples. Aspects of carbon
nanotubes can be recast into a circuit model \cite{B03,yam08} due to the
nearly found perfect electron-hole symmetry in carbon nanotube quantum dots
\cite{JSDKZ01}. One can model adiabatic charge pumping in such quantum
nanotube dots by the coupling to external leads as well as considering the 
tunneling between the dots exposed to an electric field \cite{Buit08,OP98,FBF73}. The
transport and polarization effects of quantum point contacts up to wires are
investigated with extensive engineering tools \cite{Jak06}. The dissipative
transport through such tight-binding lattices shows even a current inversion
by a nonlinearly driven field \cite{Hart97}. A high-frequency electric field
can induce artificial ferromagnetism in a tight-binding lattice \cite{VL13}. The underlying tight-binding
models are even used for modeling branched networks \cite{Buo02}.

This motivates to investigate the simpler problem of a chain of dots and their transport properties and how far it can be modeled by an equivalent circuit. Mostly quantum dots are considered with an internal quantum
structure \cite{KAT01,RRJ03,JYB06}, e.g. the coupling of quantum dots to superconducting leads \cite{ZPJN16}. For an overview about recent developments see \cite{N13}. Here we will neglect all internal features of the dots and
model simply the quantum transport of electrons through a chain of localized
dots which means we consider the confinement classically. One dimensional
scattering with confinement has been treated in \cite{ZG13}. For an overview
of one-dimensional Fermi models see \cite{V94,GBML13}. 
Here we restrict ourselves to localized dots which are essentially different
from free-moving particles like in metallic wires \cite{BMSP12} or disordered
obstacles treated usually within Landauer-B\"uttiker formalism
\cite{B86}. The localized point
contacts possess a rich magneto-transport behavior \cite{BH89} and even without disorder cold atoms in mesoscopic wires show resistivity \cite{BMSKE12}. Since mostly the continuum limit is used, we will investigate here the influence of a finite number of localized states.

We want to investigate the extensively studied tight-binding Hamiltonian with interaction and external electric fields with respect to localized positions of interacting particles. We will show that the transport through such structures can be replaced by an equivalent RCL current with resistivity, capacitance and inductance uniquely determined by a single microscopic quantum parameter. Therefore one can shape a desired transport by choosing molecules and materials for the dots according to this parameter, the latter being given in terms of
the coupling constant between the dots expressed by the nearest
neighbor hopping energy $J$, the interaction between the electrons $V$, the spacing of dots $a$ and the
effective electron density over all dots $n$.

A simple estimate allows already to get some insight into this idea \cite{B03}. The energy level spacing between quantum states in a 1D chain of length $D$ can be expressed  as 
\be
\delta E={d\epsilon_q\over dq}\delta q=2 J a \sin{(q a)} {2\pi \over D}
\label{1}
\ee
assuming a dispersion $\epsilon_q=-2 J \cos{(q a)}$ with wavelength $q$. This energy cost can be understood as realized by an effective capacitance $\delta E=e^2/C$ such that one deduces the quantum capacitance per length
\be
{C\over D}={e^2\over 2 \pi \hbar v_F}.
\label{C}
\ee
Here we introduced the "Fermi" velocity of non-interacting electrons
\be
v_F={2 J a\over \hbar }\left . \sin{(q_F a)}\right |_{q_f=\pi/2a}={2 a J\over \hbar}
\label{vf}
\ee
with the second expression valid for the quantum dots at half filling where $\epsilon_{q_F}=0$.

The kinetic inductance $L$ can be found in a similar way \cite{B03}. We
consider the potential difference $\Delta U$ between right and left
leads. The net increase of kinetic energy is the product between the
excess number of electrons in the left versus the right leads, $e \Delta U/
\delta E$, and the energy carried, $e \Delta U/2$, which provides
\be
\Delta E&=&{e^2\Delta U\over \delta E}{\Delta U\over 2}={e^2\Delta U^2 D\over 2 h
  v_f}
%\nonumber\\
= {I^2 h g^2 D\over 2 e^2 v_f}
\label{4}
\ee
where we used (\ref{1}) and that the current $I$ is given by the ratio of the potential and the
resistance 
\be
R={h\over e^2} g.
\label{R}
\ee 
All interaction effects are condensed in the
g-factor. Comparing the kinetic energy (\ref{4}) with the one expressed by the
inductance $E_{kin}=\frac 1 2 L I^2$ we deduce the kinetic inductance per 
length \cite{B03}
\be
{L\over D}={h\over e^2v_F} g^2.
\label{L}
\ee

The effective Fermi velocity $v_F^{\rm int}$ for interacting electrons is
given by the eigenfrequency $\omega_0=1/\sqrt{LC}$ times the length $D$ of the system yielding
\be
v_F^{\rm int}=\frac 1 g v_F
\ee
which is different from the non-interacting one (\ref{vf}) by the $g$-factor.

In this paper we will find the Fermi velocity as $ v_f={g \omega_0 D}$ with the g-factor
\be
g={e^2\omega_0\over h \alpha } 
\ee
where the collective frequency $ \omega_0^2=(b_{q_0}+2 n V_{q_0}) b_{q_0}$ is given in terms of the interaction potential $V_{q_0}$ and $b_{q_0}=4 J\sin^2{q_0/2}$ at the wavelength $q_0$ of the external electric field. The quantum parameter finally becomes 
\be
\alpha={n e^2\over m}{\sin{\frac {q_0}{2}}\over {q_0\over 2}}
\ee 
with the effective mass $m=\hbar^2/2 J a^2$ and lattice spacing $a$. In this way we can shape the capacitance (\ref{C}), the inductance (\ref{L}) and the resistivity (\ref{R}) by choosing appropriate materials with hopping parameter $J$, interaction $V$, density $n$ and spacing $a$ of the material.

Considering the power due to kinetic energy
\be
P=I U={d\over d t} E_{kin}=L I {d I\over dt}
\ee
shows that for a step-like switch-on of the potential $U$ the current $\dot I=U/L$ grows linearly with time. This is what has been observed in \cite{CXZCC13}.

We will investigate a chain of quantum dots now in a potential difference causing a homogeneous electric field. This will provide us with exactly this ballistic property of the current independent of interaction. This is in agreement with the observation of a transition from Ohmic to ballistic transport in quantum point contacts \cite{MBSNOEBW89}. If one uses an inhomogeneous electric field e.q. by a spatially modulated wave one obtains a nontrivial current which can be replaced by the circuit properties above.

The outline of the paper is as follows. In the next chapter we shortly review the exact expression for currents of electrons hopping between localized levels in electric fields. With the help of the exact solution of the hopping Hamiltonian  we discuss the Bloch oscillations. In the third chapter we consider interactions in mean field and relaxation-time approximation with imposed conservation laws. This chapter is divided into the decay of initial correlations and the interactions. The final results are the dynamical response function and the current. The latter one is shown to be represented by an equivalent circuit with a resistivity, capacitance and inductance found in terms of microscopic hopping, interaction and relaxation time. In the fourth chapter the pair correlation function is discussed by the analytical structure function for arbitrary temperatures.

\section{Chains in electric field}

The 1D tight-binding Hamiltonian of cites $\ket{n}$ in a time-dependent external electric field $E(t)$ reads
\be
\hat H=\sum\limits_{nn'} H(n-n')  \ket n \bra n'-e a\sum\limits_n n E_n(t)\ket n \bra n
\ee
where the position operator is
\be
\hat x&=&\sum\limits_n n a |n\rangle \langle n|
\ee
with the lattice distance $a$.
The time-dependent Schr\"odinger equation is solved by the superposition of
\be
\ket \Psi=\sum\limits_n c_n \ket n
%=\sum\limits_n \hat c_n \ket 0
\ee
where 
%we use either the operator $\hat c_n$ or 
the coefficients $c_n$ 
%which
obeys 
\be
i\hbar \partial_ tc_n=\sum\limits_{n'} H(n-n') c_{n'}-e E_n(t) a \,n\, c_n.
\label{timeS}\ee
In the following we understand the dimensionless momentum $p$ in units of $\hbar/a$ and the wave vector $k$ in units of $1/a$.
The Fourier transform according to
\be
c_p=\sum\limits_n {\rm e}^{-i p n}c_n,\quad c_n=\int\limits_0^{2\pi} {dp\over 2\pi} {\rm e}^{ip n}c_p
\label{12}
\ee
translates (\ref{timeS}) into
\be
i\hbar \partial_t c_p=\epsilon_p c_p-i e a \int \limits_0^{2 \pi} {d \bar p\over 2\pi} E_{p-\bar p}(t)  \partial_{\bar p} c_{\bar p}
\label{fieldca}
\ee
with 
\be
\epsilon_p=\sum\limits_n{\rm e}^{-ip(n-n')} H(n-n')=-2 J \cos p
\label{disp}
\ee
for nearest-neighbor hopping $H(n-n')=-J(\delta_{n',n+1}+\delta_{n',n-1})$.

A homogeneous electric field simplifies the algebra considerably and (\ref{fieldca}) reads
\be
i\hbar \partial_t c_p=\epsilon_p c_p-i e E(t) a \partial_{p} c_{p}.
\label{fieldc}
\ee
Introducing $y(p,t)=\ln{[c_p(t)]}$, this equation (\ref{fieldc}) 
can be solved exactly by an implicit representation
$F[t,p,y(p,t)]$ with
\be
F_t+eE{a\over \hbar} F_p-{i\over \hbar}\epsilon(p) F_y=0.
\label{diffF}
\ee
Since the gradient is perpendicular to any equal-potential line, the characteristics of a line in the solution plane reads
\be
1:eE{a\over \hbar}:(-{i\over \hbar}\epsilon(p))=\dot t:\dot p:\dot y
\ee
from which we obtain the two characteristics
\be
\xi_1&=&p-{a\over \hbar} \int\limits^t\!eE(\bar t) d \bar t\nonumber\\
\xi_2&=&y+{i\over \hbar}\int\limits_0^t \!d \bar t \,\epsilon\left [p+{a\over
    \hbar}\int\limits_t^{\bar t} \!eE(t') dt'\right ]
\ee
such that any function of these two constants are a solution of the partial differential equation (\ref{diffF}). Given an initial value $c_p(t=0)=c^0[p]$ the solution of (\ref{fieldc}) reads finally
\be
c_p(t)=c^0\!\!\left [p\!-\!{a\over \hbar} \int\limits_0^t\!eE(\bar t) d \bar t\! \right ] {\rm
  e}^{-{i\over \hbar} \int\limits_0^t \!d \bar t \,\epsilon\left [p+{a\over
      \hbar}\int\limits_t^{\bar t} \!eE(t') dt'\right ]}
\label{solut}
\ee
which contains all known special cases found in the literature. The constant electric field turns (\ref{solut}) into Bessel functions as obtained in \cite{FBF73}. Mean square displacements are calculated in \cite{OP98} with the intention of localizations.

With the exact solution (\ref{solut}) we can calculate the mean quantum mechanical current
\be
\dot x&=&\bra{\Psi}{i\over \hbar}[\hat H,\hat x]\ket{\Psi}=J{i\over \hbar} a\sum\limits_n(c_n^+c_{n+1}-c_{n+1}^+c_n)
\nonumber\\&=&2 J {a\over \hbar} {\rm
  Im}\sum\limits_n c_n^+c_{n+1}=2 J {a\over \hbar} {\rm
  Im}\int\limits_0^{2\pi} {dp\over 2\pi} c_p^+c_p {\rm e}^{ip}
\nonumber\\
&=&2 {a\over \hbar} J \int\limits_0^{2\pi} {dp\over 2\pi} c_p^+c_p
\sin p
\label{cur}
\ee
which is doubtlessly a real quantity since $c_p^+c_p=(c_p^+c_p)^+$.

The many-body averaging leads to the momentum-dependent density
$f_p=\langle c_p^+c_p\rangle$ with the zero center-of-mass momentum $q$. We neglect thermal effects and consider here only localized particles.
We assume dots occupying places on a lattice with a spatial distribution $n\sum_{l=-N}^N\delta(r- l a) $ which Fourier transforms to the momentum distribution
\be
f_p=n\sum\limits_{l=-N}^{N} {\rm e}^{-il p}=n {\sin{(N+1/2) {p}}\over \sin ({p/2})}.
\label{f0}
\ee

With the help of the exact solution (\ref{solut}) and abbreviating ${\cal E}=
{a\over \hbar}\int\limits_0^t\!e E(t')dt'$ one calculates the current (\ref{cur})
as
\be
\langle \dot x\rangle&=&2 J {a\over \hbar} \int\limits_{{\cal E}}^{2\pi +{\cal
    E}}\sin{(p+{\cal E})}f_p=
2 J {a\over \hbar} \int\limits_{0}^{2\pi}\sin{(p+{\cal E})}f_p\nonumber\\
&=&2 J {a\over \hbar}  n \sin{{\cal E}}=2 J {a\over \hbar}  n \sin{\left (
{a\over \hbar}\int\limits_0^t\!e E(t')dt'\right )}.
\label{curr}
\ee
In the second step, we have used the Brillouin-zone ($2\pi$)-periodicity of $f_p$ and observe
that its integral over $\sin{p}$ is zero and the $\cos{p}$-weighted integral leads
to the same value as the normalization itself. 

As a consequence, one has inevitable Bloch oscillations which for
constant fields takes the known form 
\be
 \langle \dot x\rangle&=&2 n J {a\over \hbar}   \sin{\left (e
{a\over \hbar} E t\right )}.
\ee
In linear response (\ref{curr}) leads to 
\be
\langle \dot x \rangle&=&2 n J {a^2\over \hbar^2}   \int\limits_0^t
E(t')dt'={n\over m}\int\limits_0^t
E(t')dt'
\label{currlin}
\ee
where the effective mass near the band edge has been used according to
$\epsilon_p=-2 J \cos(p a/\hbar)\approx -2 J+J{a^2\over \hbar^2}
p^2$. Eq. (\ref{currlin}) describes nothing but the free
ballistic motion in a time-varying homogeneous electric field. In other words the chain of
quantum dots with only hopping between neighbors leads to ballistic currents.

\section{Currents and response}

\subsection{Hopping and decay of initial correlations}
Let us inspect how this situation changes if we add interactions and consider the linear response. First we
reformulate the hopping situation within the kinetic theory
and then we investigate the interactions. 
Therefore we search for the Wigner function $\rho_{1,2}=\bra 1 \hat \rho \ket 2$ and represent the Heisenberg equation
\be
i\hbar \dot \hat \rho=[\hat H,\hat \rho]
\label{heisen}
\ee
in matrix form for the interaction-free case,
\be
i\hbar \dot \rho_{1,2}=(\epsilon_1-\epsilon_2) \rho_{1,2}+\sum\limits_3( U_{1-3}\rho_{3,2}-U_{3-2}\rho_{1,3})
\label{h1}
\ee
where $U$ is the external potential. In linear response around the equilibrium
$\rho_{1,2}=f_1\delta_{1,2}$ and using momentum representation $\bra 1=\bra {p+q/2}$ and $\bra 2=\bra
{p-q/2}$ one gets for (\ref{h1})
\be
i\hbar \dot {\delta \rho_{pq}}=\Delta \epsilon \delta \rho_{pq}+U_q \Delta f
\label{eq}
\ee
with 
\be
&&\Delta \epsilon=\epsilon_{p+q/2}-\epsilon_{p-q/2}=4 J \sin{(q/2)} \sin p\nonumber\\
&&\Delta f=f_{p+q/2}-f_{p-q/2}.
\label{not}
\ee
Eq. (\ref{eq}) is readily solved as
\be
\delta \rho_{pq}&=&\rho_q(0){\rm e}^{-{i\over \hbar}\Delta \epsilon t}+{i\over \hbar}\int\limits_0^td\bar t U_q(t-\bar t) {\rm e}^{-{i\over \hbar}\Delta \epsilon \bar t}\Delta f\nonumber\\
&=&\rho_q(0){\rm e}^{-{i\over \hbar}\Delta \epsilon t}+{\Delta f\over
  \Delta \epsilon} \left (1-{\rm e}^{-{i\over \hbar}\Delta \epsilon t}\right )U_q
\label{sol}
\ee
with the second line for time-independent external potentials.

For large total number of dots $N$ which we will consider first, one gets from
(\ref{f0}) 
\ba
f_p=&\lim\limits_{N\to \infty}\!n {\sin{(N\!+\!\frac 1 2) {p}}\over \sin ({p\over2})}=
\lim\limits_{N\to \infty}\!n {\sin{(N\!+\!\frac 1 2) {p}}\over {p\over 2}}{{p\over 2}\over\sin ({p\over 2})}\nonumber\\
&=\lim\limits_{N\to \infty}n 2 \pi  n\delta (p) {{p\over 2}\over\sin ({p\over 2})}=2 \pi  n\delta (p)
\label{f}
\end{align} 
for the interval $p\in(0,2\pi)$ and $f_p$ are $2\pi$ periodic. We can consider this as a model distribution for completely momentum-localized states like in Bose-Einstein condensation.
This renders the momentum integration trivial.
One obtains from (\ref{sol}) for the density 
\be
\delta n_q(t)&=&\rho_q(0) J_0\left [4 J t \sin\left (\frac q 2\right )\right
]\nonumber\\
&&-2 n \int\limits_0^td\bar t U_q(t-\bar t)\sin{\left [4 J \bar t
  \sin^2\left (\frac q 2\right ) \right ]}
\label{dn}
\ee
with the Bessel function $J_0$.
The first part gives the decay of initial density and the second part the change due to the external potential. The decay of initial correlations is present even without an external potential and any interaction. They decay due to the Bessel function with $1/\sqrt{t}$. This is a result of quantum interference as it will be expressed by the associated current density which turns out to be purely imaginary.

\begin{figure}[]
\includegraphics[width=7cm]{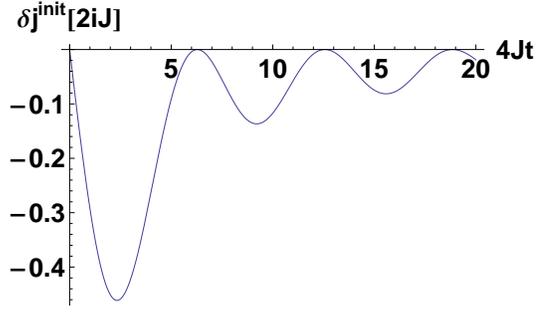}
\caption{\label{imagceps}The imaginary current (\ref{jcinit}) integrated over the Brillouin zone wave vectors which is connected to the decay of initial correlations.}
\end{figure}

The total current according to (\ref{cur}) is
\be
\delta j(t)&=&2{a\over \hbar} J \int \limits_{-\infty}^{\infty} {d q\over 2\pi}\int \limits_0^{2\pi} {d p\over 2\pi} \delta \rho_{pq}(t) \sin p=\int \limits_{-\infty}^{\infty} {d q\over 2\pi}j_q(t)
\nonumber\\
j_q&=&{a\over \hbar}\int\limits_0^{2\pi}{d p\over 2\pi} {\partial_p \epsilon_p}\delta \rho_{pq}.
%&=&-2 J a \int \limits_0^{2\pi} {d q\over 2\pi} \delta n_{-q-2\pi} (t) \sin q
\label{jj}
\ee
%where the second relation is a simple substitution.
It is convenient to discuss the initial and potential parts separately.
The first part of (\ref{sol}) leads to no real current density as one can see from
\be
j_q^{\rm init}(t)&=&2 J {a\over \hbar}\int\limits_0^{2\pi} {d p\over 2 \pi} \sin(p) \rho_q(0){\rm
  e}^{-ix \sin p}
\nonumber\\
&=&2 i {a\over \hbar} J \rho_q(0) \partial_{x}   \int\limits_0^{2\pi} {d p\over 2 \pi} {\rm
  e}^{-ix \sin p}
\nonumber\\
&=&
2 i {a\over \hbar}  J \rho_q(0) J_0'[x]=-2 i {a\over \hbar}  J \rho_q(0) J_1[x]
\label{jcinit}
\ee
with $x=\sqrt{4 J b} t$ and $b=4 J \sin^2(q/2)$. 
 This result means that the initial distribution decays with an imaginary current
density and the Brillouin-zone-integrated imaginary current $(0,2\pi)$ 
is plotted in the figure~\ref{imagceps}. 

Though the initial density
decays with $1/\sqrt{t}$ in (\ref{dn}) it is not connected with a real
current. Indeed, the total integrated current density for $q\in (-\infty,\infty)$ is zero. Therefore we interpret this as quantum correlation decay. 

In contrast, the second part of (\ref{sol}) in (\ref{jj}) due to the external perturbation will lead to a real
current. Lets assume a time-varying electric field and its
potential in one dimension being $U(x)=-e \int\limits_0^xE_{x'}(t) dx'$. A homogeneous electric field in a chain of length $D$ has the potential
\ba
U_q\!&=\!\!\lim\limits_{D\to\infty} a\!\! \int\limits_{-D/2a}^{D/2a}
\!\!d{x} {\rm e}^{-i q x} [-e E(t)x]
\nonumber\\
&=-2i a e E(t)\partial_q \lim\limits_{D\to\infty}\!{\sin{q\frac{D}{2a}}\over
    q}
=-i 2 \pi a e E(t)\partial_q \delta(q) 
\label{Uq}
\end{align}
while an inhomogeneous electric field with wavelength $q/\hbar$ possesses the potential
\be
U_q= e a i {E_q(t)\over q}.
\label{Uqq}
\ee

Using the homogeneous potential (\ref{Uq}) in the second part of (\ref{sol}), the current (\ref{jj})
is easily evaluated. Since we are interested in the current at position $x=0$ we integrate over $q$ and obtain
\ba
& 4 {a^2\over \hbar} J \!\!\int\limits_{-\infty}^{\infty}\!\! {d
  q}\delta'(q)\!\!\int\limits_0^{2\pi}\!\! {d p\over 2\pi}\sin
p\!\!\int\limits_0^t\!\!d\bar t \,e E(\bar t) \Delta f {\rm e}^{-{i\over \hbar} 4
  (t\!-\!\bar t) J \!\sin{\!\frac q 2} \sin{\!p}} 
\nonumber\\
&=- 2 {a^2\over \hbar} J\! \int\limits_{-\infty}^{\infty}\! {d
  q}\delta(q)\!\int\limits_0^{2\pi} \!{d p\over 2\pi}\sin
p\!\int\limits_0^t\!d\bar t e E(\bar t) \partial_p\left (f_{p\!+\!\frac q 2}\!+\!f_{p\!-\!\frac q 2}\right )
\nonumber\\
&=2 {a^2\over \hbar} J \int\limits_0^{2\pi} {d p\over 2\pi} f_p\cos p
\int\limits_0^td\bar t e E(\bar t)%\nonumber\\ &
=2 {a^2\over \hbar} J n \int\limits_0^td\bar t e E(\bar t)
\end{align}
where twice partial integrations have been performed and we note that the $\cos$-weighted density equals the density $n$ for the distribution (\ref{f0}). 

We obtain again the result that a chain of quantum dots in a homogeneous field with hopping in-between the dots will
lead to a ballistic current as it was observed in \cite{CXZCC13}. This is valid for finite and infinite chains. It is not hard to see that even a finite-length-D potential in (\ref{Uq}) does not alter this result.

\subsection{Hopping and interaction}

Next we look into interacting electrons hopping in a chain of dots. Therefore we consider the Hamiltonian
\be
\hat H=\sum\limits_p\epsilon_p \hat c_p^+\hat c_p+\frac 1 2 \sum\limits_{p,k,q} V_q
\hat c_{p+q}^+\hat c_{k-q}^+\hat c_k\hat c_p
\ee
which has the dispersion (\ref{disp}) and the electrons interact with the potential $V_q$.
The kinetic equation is written again with the help of (\ref{heisen}) in momentum representation analogously to (\ref{eq}) but now with interaction $V_q$. Its 
Laplace transform reads \cite{M14a}
\ba
(s+\frac 1 \tau +i\Delta \epsilon )\delta f-\delta f_0=i\Delta f (U_q+\delta n V_q)+{\Delta f\over \Delta \epsilon} {\delta n\over \tau \int\limits_0^{2\pi}\!\!{d p\over 2 \pi} {\Delta f\over \Delta \epsilon}} 
\end{align}
with the meanfield approximation and a density-conserving relaxation-time approximation a l\'a Mermin \cite{Mer70,D75}. It can be extended to include more conservation laws \cite{Ms01,Ms12}. The notation of (\ref{not}) is used.
Solving and integrating over $p$, the density change due to the external field $U_q$ reads 
\be
\delta n_q= {\Pi_S U_q-i \delta f_0 Q_S\over 1-V_q \Pi_S -{1\over \tau S} \left (1-{\Pi_S\over \Pi_0}\right )}
\label{res}
\ee
where we use $S=s+\frac 1 \tau$. We have introduced the RPA polarization and initial polarization
\ba
\Pi_S&=\int\limits_0^{2\pi} {d p\over 2\pi} {\Delta f\over \Delta \epsilon-i S} \nonumber\\
Q_S&=\!\int\limits_0^{2\pi}\!\! {d p\over 2\pi} {1\over \Delta \epsilon\!-\!i S}=\!{i\over \sqrt{S^2\!+\!4Jb}} \,\,\Laplace \,\, i J_0(\sqrt{4 J b}t){\rm e}^{-{t\over\tau}}
\label{pol}
\end{align}
with $b=4 J\sin^2{q/2}$. We gave the Laplace back transformation into time of the initial polarization in terms of the Bessel function $J_0$.
The corresponding current density according to (\ref{cur}) reads 
\be
\delta j_q= \delta j_q^{\rm init}+\delta j_q^{c}
\ee
with the current due to initial correlations 
\ba
\delta j_q^{\rm init}&=-2i J {a\over \hbar} \delta f_0
\!\left [
Q^s_S\!+\!Q\Pi_S^s{V_q \!-\!{1\over \tau S\Pi_0}\over
1\!-\!V_q \Pi_S \!-\!{1\over \tau S} \left (1\!-\!{\Pi_S\over \Pi_0}\right )
}
\right ]
\label{jinit}
\end{align}
and the modified polarization function
\be
\Pi_S^s&=&\int\limits_0^{2\pi} {d p\over 2\pi} \sin{p} {\Delta f\over \Delta \epsilon-i S} \nonumber\\
Q_S^s&=&\int\limits_0^{2\pi} {d p\over 2\pi} {\sin{p} \over \Delta \epsilon-i S}={\sqrt{4 J b}\over \sqrt{S^2+4Jb}(\sqrt{S^2+4Jb}+S)}\nonumber\\
&\Laplace& J_1(\sqrt{4 Jb} t){\rm e}^{-{t\over\tau}}.
\label{polmod}
\ee
Note that $\Pi^s_0=0$ due to symmetries.
We see that the first part of (\ref{jinit}) is just the free decay of initial correlations as we had seen in (\ref{jcinit}). 

The current due to the external potential $U_q$ reads now
\ba
\delta j_q^{\rm c}&=&2 J {a\over \hbar} U_q\left [
\Pi^s_S+\Pi\,\Pi_S^s{V_q -{1\over \tau S\Pi_0}\over
1-V_q \Pi_S -{1\over \tau S} \left (1-{\Pi_S\over \Pi_0}\right )
}
\right ].
\label{jc}
\end{align}

It is easy to integrate over $q$ to get the total current for a homogeneous electric field with
the potential (\ref{Uq}) for the dispersion (\ref{disp}).  With the help of 
\be
\left . {\Delta f\over \Delta \epsilon}\right |_{q=0}={f'_p\over 2 J \sin{p}};\qquad \left .\partial_q
 {\Delta f\over \Delta \epsilon}\right |_{q=0}=0
\ee
the current due to the homogeneous external field is just the free ballistic one
\be
\delta j_q^{\rm c}=2 {a^2\over \hbar} J eE(s)  n \frac{1}{S}\,\Laplace \, 2 {a^2\over \hbar} J \int\limits_0^t d\bar t e E_{t-\bar t} {\rm e}^{-{\bar t\over\tau}} 
\ee
as obtained for non-interacting chains except the folding with the relaxation which describes exactly the dissipative decay of the current. 

Further we want to consider the case of large chains which allows to use (\ref{f}) with the help of which the polarizations (\ref{pol}) and (\ref{polmod}) take the simple forms
\be
\Pi_S&=&-2 n {b\over S^2+b^2}\nonumber\\  
\Pi^s_S&=&-2 i n \sin{\left ({q\over 2}\right )}{S\over S^2+b^2}  
\label{simple}
\ee
and the currents (\ref{jinit}) and (\ref{jc}) become
\ba
\delta j_q^{\rm init}&=-2i J {a\over \hbar} \delta f_0
\left [
{\sqrt{4 J b}\over \sqrt{S^2+4Jb}(\sqrt{S^2+4Jb}+S)}
\right .\nonumber\\
&
\left .+{\sin\left ({\frac q 2}\right )\over \sqrt{S^2+4J b}}{2 n V_q S+ \frac  b \tau \over b^2+S^2+2 n V_q b -\frac S \tau}
\right ]
\end{align}
and
\ba
\delta j_q^{\rm c}&=-4i J {a\over \hbar} n E_q \sin\left ({\frac q 2}\right )
\left [{S\over b^2+S^2}
\right .\nonumber\\
&\left . -
{b\over S^2+b^2}{2 n V_q S+ \frac  b \tau \over b^2+S^2+2 n V_q b -\frac S \tau}
\right ].
\label{49}
\end{align}
The currents Laplace-transform back into time
\ba
&\delta j_q^{\rm init}=-2i J {a\over \hbar} \delta f_0 \left [
J_1(\sqrt{4 Jb} t) {\rm e}^{-{t\over \tau}}
+  \sin{\frac q 2} {\rm e}^{-{\frac t \tau}}
\right .\nonumber\\
&\left .\times \!\int\limits_0^t\!\!d {\bar t} J_0(\sqrt{4J b} (t\!-\!{\bar t}))(2 n V_q \cos{\gamma {\bar t}}\!+\!(n V_q\!-\!b) {\sin{\gamma {\bar t}}\over 2 \gamma \tau}){\rm e}^{{{\bar t}\over 2\tau}}
\right ]
\end{align}
and
\ba
&\delta j_q^{\rm c}=-4i J {a\over \hbar} n U_q \sin{\frac q 2}  
{\rm e}^{-\frac{t}{2\tau}}
\left (\cos{\gamma_q {t}}-{\sin{\gamma_q {t}}\over 2 \gamma_q \tau}\right )
\end{align}
with $\gamma^2_q=b_q^2+2 n V_q b_q-1/4\tau^2$. 

It is now interesting to investigate the case of an inhomogeneous field (\ref{Uqq}) which we consider for a single-mode wavelength
\be
U_q=2\pi i E_{q_0} {a\over q_0}\delta (q-q_0).
\label{onem}
\ee
This means we can replace the wavelength $q$ in (\ref{jc}) just by $q_0$ and divide the result by $q_0$. 
For the one-mode electric field with the potential (\ref{onem}) one can integrate the current density over wave vectors and multiply with $e$ to get the total charge current. We represent it in terms of the resistance in frequency space which means we replace in the Laplace transform of (\ref{49}) remembering $S=-i\omega+1/\tau$ and $\delta J^c=\delta j_{q=0}^c$ to get
\ba
&R^c={E_{q_0}\over \delta J^{\rm c}} =
{
\frac 1 \tau +i\left ({b_{q_0}(b_{q_0}+2 n V_{q_0})\over \omega}-\omega\right ) 
\over
\alpha}
\label{romega}
\end{align}
with 
\be
\alpha=2e^2 J {a^2 n\over \hbar^2}{\sin{\frac {q_0}{2}}\over q_0/2}={n e^2\over m}{\sin{\frac {q_0}{2}}\over {q_0\over 2}}
\label{alpha}
\ee
for the effective mass for free particles $m=\hbar^2/2 J a^2$.

We can equivalently replace the chain of quantum dots by a damped oscillator circuit with the Ohmic resistance per length
\be
R^{-1}= \tau \alpha={n e^2 \tau\over m}{\sin{\frac {q_0}{2}}\over {q_0\over 2}}.
\ee
The finite damping leads to an inverse resistivity $R$ per length in 1D which is equivalent to the known conductivity except the modulation factor due to the applied wave ${\sin{q_0/2}\over q_0/2}$.

\begin{figure}[]
\includegraphics[width=8cm]{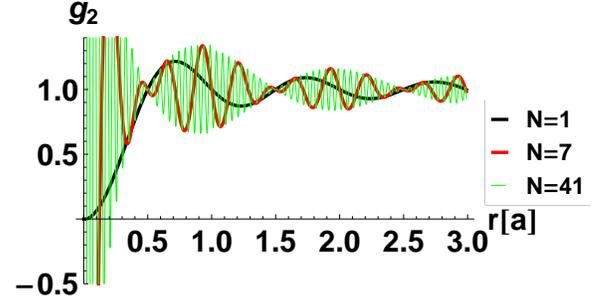}
\caption{Pair correlation function (\ref{gr}) for different sites, a hopping strengths of $J=0.001 T$, interaction $V=1 a T$, and relaxation time $\tau =0.3\hbar /T$, each in units of temperature $T$.\label{realg_j001_v1_tt03_t1}}
\end{figure}

From (\ref{romega}) we read-off also the equivalent inductance $R_L=-i \omega L$ per length
\be
L^{-1}=\alpha
\ee
and the equivalent capacitance $R_C=i/\omega C$ per length
\be
C={\alpha\over \omega_0^2}.
\ee
Here we have introduced the eigenfrequency for undamped electrons $\tau\to\infty$
\be
\omega_0^2={1\over L C}=(b_{q_0}+2 n V_{q_0}) b_{q_0}.
\label{eigen}
\ee

In order to make the connection to the introduction we can use the replacement
\be
g={e^2 \omega_0\over 2\pi \hbar \alpha } ;\qquad v_f={g \omega_0 D}={e^2
  \omega_0^2 D\over 2\pi \hbar  \alpha }
\label{gv}
\ee
to get the kinetic inductance and quantum capacitance.  The equations (\ref{alpha})-(\ref{gv}) are the main results of this paper. We have derived the expression of the g-factor for the interacting case which results into the determination of all equivalent RCL circuit measures by a single microscopic parameter (\ref{alpha}).

\section{Pair correlation function}

\begin{figure}[]
\includegraphics[width=8cm]{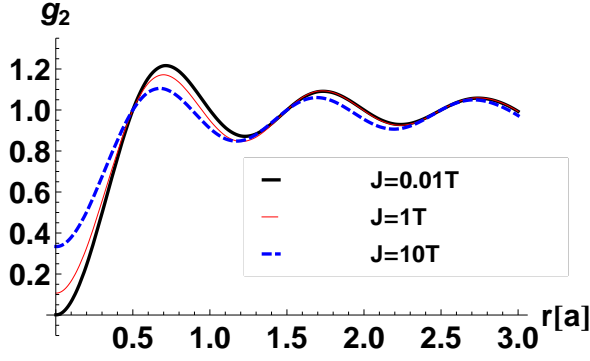}
\caption{Pair correlation function (\ref{gr}) of one Brillouin zone for different hopping strengths and $V=1 a T$, $\tau =0.3\hbar /T$.\label{realg_v1_tt03_t1}}
\end{figure}

It is also possible to give an analytic expression for the structure factor
when we consider the case of large chains which had already resulted into the simple distribution (\ref{f}) and the response functions (\ref{pol}). The finite-temperature structure factor is given in terms of the response function $\chi=\delta n_q/U_q$ as
\be
S_q=-{1\over n}\int\limits_{-\infty}^\infty {d \omega\over \pi} {{\rm Im} \chi\over 1-{\rm e}^{-\beta \omega}}
\ee
with $\beta=1/T$. We consider the temperature effects in the response function as the temperature of the surrounding bath though we have considered perfectly localized dots without thermal motion. Using (\ref{simple}) in (\ref{res}) one obtains without initial correlations
\be
S_q&=&{\rm Im}{i b_q\over \pi}\int\limits_{{1\over \tau}-i\infty}^{{1\over \tau}+i\infty} \!\!d s{1\over (1\!-\!{\rm e}^{-i\beta (s\!-\!\frac 1 \tau)})(b_q^2\!+\!n b_q V\!+\!s^2\!-\!\frac s \tau)}
\nonumber\\
&=&{\rm Im} {b_q\over 2 \pi \gamma_q} \left [ \Psi\left (-{\beta \over 4 \pi \tau}\!+\!{i \beta \gamma_q\over 2 \pi}\right )\!-\!\Psi\left (-{\beta \over 4 \pi \tau}\!-\!{i\beta \gamma_q\over 2 \pi}\right )
\right .\nonumber\\&& \left .
-2 \pi i {\sinh{\beta \gamma_q}\over \cosh{\beta \gamma_q}-\cos{\beta \over 2 \tau}} \right ]  
\label{Sq}
\ee 
with the DiGamma function $\Psi(x)=\Gamma'/\Gamma$ coming from the residue of
the Bose function and the remaining parts from the residua of the quadratic
denominator. We have used $\gamma_q^2=b_q^2+2 n V b_q-1/4\tau^2$ and if
$\gamma_q^2<0$ one has to use $\gamma_q=i \bar \gamma_q$. The zero-temperature
result is analytically as well and reads
\be
S_q(T=0)={\rm Im}{b_q\over 2 \pi \gamma_q} \ln{1+2 i\gamma_q\tau \over 1-2i\gamma_q \tau}.
\ee

\begin{figure}[]
\includegraphics[width=8cm]{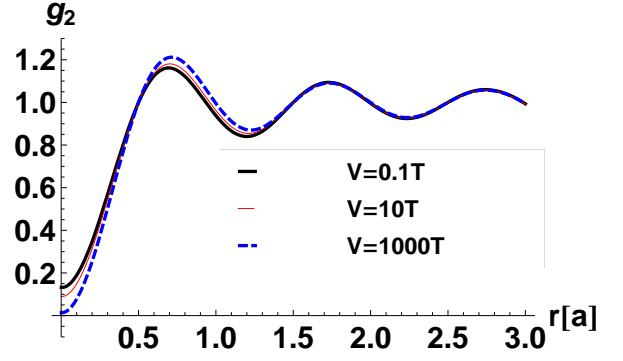}
\caption{Pair correlation function (\ref{gr}) for different interactions and $J=1 T$, $\tau =1\hbar /T$.\label{realg_j1_tt1_t1}}
\end{figure}

The pair correlation function is then given by
\be
g_r&=&1+{1\over 2 \pi n a}\int\limits_{-\infty}^{\infty} dq {\rm e}^{i q r} (S_q-1)
\nonumber\\
&=&1+{1\over 2 \pi n a}\int\limits_{-\infty}^{\infty} dq \cos{(q r)} (S_q-1)
\nonumber\\
&=&1+{1\over 2 \pi n a}F_N(r)
%\nonumber\\&&\qquad \times 
\int\limits_{0}^{2\pi} dq \cos{(q r)} (S_q-1)
\label{gr}
\ee
where we used the periodicity of $S_q$ and integrate over $2N+1$ sites which results into a structure factor $F_N(r)=\sin{[\pi r (2N+1)]}/\sin{[\pi r]}$ in front of the integration over one Brillouin zone. 
%We understood $q$ in units of $\hbar/a$ and the spatial distance in units of $a$. 
The structure factor is $a$-periodic $F_N(r)=F_ N(r+a)$ with maxims at $r=n a$ and the main maxim at $F_N(0)=2 N+1$. We present the pair correlation function in figure \ref{realg_j001_v1_tt03_t1} for different length of the wire. One sees that the effect of a larger number of dots is a modulation of the first-Brillouin-zone result with maximal modulation amplitude at the dot position $n a$. Interestingly every time when the first-Brillouin-zone pair correlation function crosses unity, no modulation appears. In the following we restrict to one site $N=1$ yielding the pair-correlation function of one Brillouin zone or form factor.

\begin{figure}[]
\includegraphics[width=8cm]{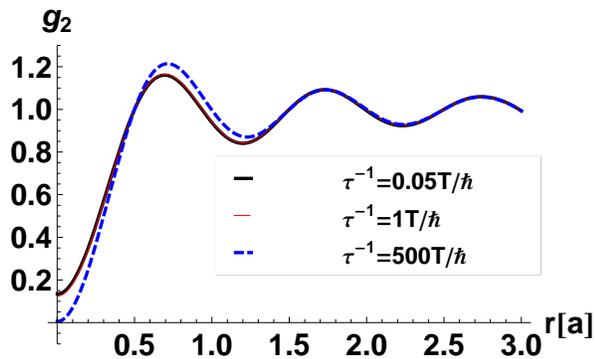}
\caption{Pair correlation function (\ref{gr}) for different relaxation times and $J=1 T$, $V =1 a T$.\label{realg_j1_v1_t1}}
\end{figure}

In figure \ref{realg_v1_tt03_t1} we plot the dependence of the pair correlation function from one Brillouin zone on the hopping strength. We see that the typical maxims indicating the distance of the effective nearest neighbors slightly change with the hopping strength. A larger hopping strength leads to a weaker correlation hole at shorter distances and the pair correlation becomes smoothed out. It counteracts the correlations. 
%At zero distance the pair correlation becomes negative for strong hopping which is a hint to instability. 
As an artifact of the RPA approximation, the pair correlation function may become negative as can be seen in metallic wires \cite{BMSP12}.
 
The dependence on the interaction determining the collective mode (\ref{eigen})  and on the relaxation times is much weaker as demonstrated in the next figure \ref{realg_j1_tt1_t1} and \ref{realg_j1_v1_t1}. The effect of interaction due to relaxation time as well as interaction affects the pair correlation oppositely than the hopping. A higher collision frequency and a higher interaction both lead to sharper features in the pair correlation. Therefore we see the expected result that the dissipative interaction represented by the relaxation times and the hopping strength both counteract in the pair correlation and the interaction leads to a deeper correlation hole.

\begin{figure}[]
\includegraphics[width=8cm]{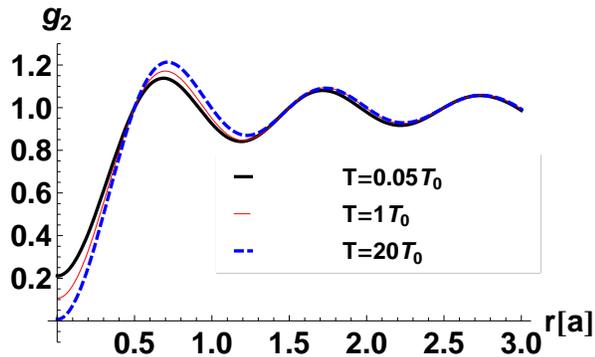}
\caption{Pair correlation function (\ref{gr}) for different temperatures and $J=1 T_0$, $V =1 a T_0$ and $\tau= 0.3\hbar/T_0$.\label{realg_j1_v1_tt03}}
\end{figure}

The dependence on the temperature finally is seen in figure \ref{realg_j1_v1_tt03}. Lower temperatures have the same effect as stronger hopping and lower correlations.

\section{Summary}

We have considered localized electrons in electric fields allowing for hopping and interactions. The exact analytical expressions show Bloch oscillations and ballistic transport for non-interacting electrons allowing only hopping between dots. Including interactions the corresponding kinetic equation can be solved in linear response and the currents are calculated analytically. We find that the transport of electrons in a chain of such dots can be represented by an equivalent R-C-L circuit. We derive explicit expressions for the equivalent resistivity, conductance and inductance in terms of hopping, interaction strength and relaxation time. The decay of initial correlations is realized by virtual currents. The pair correlation function is discussed due to the analytic expression for the dynamical response function. Here the effect of hopping counter-acts the effect of interactions and collisions. A higher temperature sharpens the feature of the first Brillouin zone. The number of dots leads to a structure factor inside the pair-correlation function which modulates the first-Brillouin-zone correlation function away from the points crossing unity. This modulation has maxims at the places of the dots.
%\appendix
%\section{Tight-binding in external field}
%
%
%\subsection{Constant field and Wannier-Stark levels}
%Following closely the work of \cite{FBF73} first lets recall the results for the infinite chain and then the finite chain.
%
%\subsection{TIme-dependent fields and localization} 

%\bibliography{bose,kmsr,kmsr1,kmsr2,kmsr3,kmsr4,kmsr5,kmsr6,kmsr7,delay2,spin,spin1,refer,delay3,gdr,chaos,sem3,sem1,sem2,short,cauchy,genn,paradox,deform,shuttling,blase,spinhall,spincurrent,tdgl,pattern,zitter,isospin,quench,hubbard}
%\bibliographystyle{prsty}

\end{document}